\documentclass[vecphys]{svmult}		
\usepackage{makeidx}         
\usepackage{epsfig}          
\begin{document}

\title*{Beyond Gross-Pitaevskii Mean Field Theory}
%
\author{
N.P.~Proukakis
}

\institute{
School of Mathematics and Statistics, Newcastle University,
Merz Court, Newcastle-upon-Tyne, NE1 7RU, United Kingdom
\texttt{Nikolaos.Proukakis@ncl.ac.uk}
}
\maketitle


\section{Introduction}

A large number of effects related to the phenomenon of Bose-Einstein Condensation (BEC)
can be understood in terms of lowest order mean field theory, whereby the 
entire system is assumed to be condensed, with thermal and quantum fluctuations completely
ignored \cite{Pethick_Book}.
Such a treatment leads to the Gross-Pitaevskii Equation (GPE) \cite{Gross}
used extensively throughout this book. Although this theory works
remarkably well for a broad range of experimental parameters, a more complete treatment is
required for understanding various experiments, including experiments with solitons and vortices. Such treatments should include the 
dynamical coupling of the condensate to the
thermal cloud, the effect of dimensionality, the role of quantum
fluctuations, and should also describe the critical regime, including the process of condensate formation.

The aim of this Chapter is to give a brief but insightful overview of various recent theories, which extend beyond the GPE (some of which may
have already been mentioned in earlier chapters).
As the author has been heavily involved in the development and implementation of such theories, the presentation will follow a somewhat personal perspective, following a logical progression from simple to more complex approaches, rather than a historical account. To keep the discussion brief,
only the main notions and conclusions will be presented. While numerous
appropriate key references are given, it is unfortunately not possible to 
mention all works that have contributed to the vast and growing field of finite temperature theories,
and the author apologizes in advance for any such potential omissions.

This Chapter generalizes the presentation of Chapter 1, by explicitly maintaining fluctuations around the condensate order parameter.
While the theoretical arguments outlined here are generic, the emphasis is on approaches  suitable for describing {\em single weakly-interacting atomic} Bose gases in {\em harmonic} traps. Interesting effects arising when condensates are trapped in double-well potentials  and optical lattices, as well as the cases of spinor condensates, and atomic-molecular coupling, along with the modified or alternative theories needed to describe them, will not be covered here.


\section{Microscopic Derivation of the Gross-Pitaevskii Equation \label{GPE_Derivation}}

For a sufficiently dilute ultracold atomic gas, where 
only binary interactions are relevant,
the Hamiltonian of the system takes the form
\begin{eqnarray}
{\hat H} =  \int  d{\bf r} {\hat \Psi}^{\dagger}({\bf r},t) 
\hat{H}_{0} {\hat \Psi}({\bf r},t) 
+ \frac{1}{2} \int d{\bf r} d{\bf r'} {\hat \Psi}^{\dagger}({\bf r},t) 
{\hat \Psi}^{\dagger}({\bf r'},t) V({\bf r}-{\bf r}'){\hat \Psi}({\bf r}',t) {\hat \Psi}({\bf r},t) 
\label{H}
\end{eqnarray}
where ${\hat \Psi}^{(\dagger)}({\bf r},t)$ are the Bose field annihilation (creation)
operators, $V({\bf r}-{\bf r}')$ is the two-body interatomic potential, and
$\hat{H}_{0}=-(\hbar^2/2m) \nabla^2 + V_{\rm ext}({\bf r},t)$ is the `single-particle' operator in an external potential $V_{\rm ext}({\bf r},t)$.
In the usual discussion of BEC, the Bose field operator 
is typically split into two parts \cite{Fetter_1972}
\begin{equation}
{\hat \Psi}^{\left( \dagger \right)}({\bf r}, t) = \Psi^{(*)} ({\bf r}, t) \hat{\zeta}^{\left( \dagger \right) }({\bf r},t) + {\hat \psi}'^{\left( \dagger \right) }({\bf r}, t) \label{zet},
\end{equation}
respectively corresponding to the condensate, and non-condensate contributions.
Here $\Psi^{(*)}$ is the complex `classical' amplitude of the condensate, while the operator 
$\hat{\zeta}^{\left( \dagger \right)} $
is related to the annihilation (creation) of a particle from the condensate and
accounts for fluctuations in the condensate phase.
%
Initially, we restrict ourselves to relatively large three-dimensional (3D) systems, sufficiently far from the regime of
critical fluctuations, where we can, to good approximation,
ignore the effect of the condensate operator $\hat{\zeta}^{(\dag)}$.
This approximation relies on the notion of spontaneous breaking of the $U(1)$ symmetry, and leads to the appearance of a condensate mean field, $\Psi({\bf r}, t)$, often termed the `condensate wavefunction', with all related fluctuations contained in the `non-condensate' operator $\hat{\psi}'({\bf r}, t)$. We will return to the role and physical interpretation of $\hat{\zeta}^{(\dag)}({\bf r},t)$
at appropriate points in this Chapter.\\

\noindent {\em Symmetry-Breaking Picture:}
We expand the Bose field operator in terms of a complete basis set $\{ \phi_{i}({\bf r}) \}$ via 
$\hat{\Psi}^{\left( \dagger \right)}({\bf r},t) = \sum_{i} \hat{a}_{i}^{\left( \dagger \right)}(t) \phi_{i}^{(*)}({\bf r})$, where brackets distinguish between creation and annihilation operators.
We assume that each single-particle mode occupied by the condensate can be expressed in terms of a mean-field amplitude $z_{i}^{(*)}=\langle \hat{a}_{i}^{ ( \dag )} \rangle$, and a fluctuation operator
$\hat{c}_{i}^{( \dag ) } = \hat{a}_{i}^{ ( \dag )}- \langle \hat{a}_{i}^{ ( \dag )} \rangle$, obeying $\langle \hat{c}_{i}^{( \dag ) } \rangle = 0$, where $\langle \cdots \rangle$ denotes averaging over all states.
The {\em exact} equation of motion for the condensate mean field amplitude of a general state $n$ thus becomes \cite{Proukakis_Thesis,Proukakis_NIST}
\begin{equation}
i \hbar \frac{d z_{n}}{dt} = \hbar \omega_{n} z_{n} + \sum_{ijk} \langle ni | \hat{V} | jk \rangle 
\left[ z_{i}^{*} z_{j} z_{k} + 2 \langle \hat{c}_{i}^{\dag} \hat{c}_{j} \rangle z_{k} 
+ z_{i}^{*} \langle \hat{c}_{j} \hat{c}_{k} \rangle + \langle \hat{c}_{i}^{\dag} \hat{c}_{j} \hat{c}_{k} \rangle \right], \label{zn}
\end{equation} 
where $\langle ni | \hat{V} | jk \rangle   = (1/2) \{ ( ni| \hat{V} |  jk ) +   ( ni| \hat{V} |  kj ) \}$ is the interaction matrix element with
$( ni| \hat{V} |  jk ) = \int \int d{\bf r}d{\bf r^{'}} \psi_{n}^{*}({\bf r})  \psi_{i}^{*}({\bf r^{'}})  V({\bf r} - {\bf r^{'}}) \psi_{j}({\bf r^{'}})  \psi_{k}({\bf r}) $, and
$\hbar \omega_{n} = \langle n | \hat{H}_{0} | n \rangle$.
The first three terms in the sum of Eq. ({\ref{zn}) depend on the condensate mean field, 
with the second term additionally depending on the 
non-condensate density, 
$\langle \hat{c}_{i}^{\dag} \hat{c}_{j} \rangle$, and
the third term on the average of two annihilation operators, 
$\langle \hat{c}_{j} \hat{c}_{k} \rangle$, termed the `anomalous average'. 
Importantly, however, Eq. (3) contains an additional term due to correlations of {\em three} fluctuation operators which does {\em not} depend directly on the condensate mean field.
Although such `triplet' terms $\langle \hat{c}_{i}^{\dag} \hat{c}_{j} \hat{c}_{k} \rangle$ were ignored
in the majority of early treatments
(see, e.g. \cite{Giorgini_Linear_Response_1,ZNG_PRL_1,Griffin_HFB} and references therein), 
it is {\em precisely} these 
contributions, which enable the transfer of atoms and energy between the condensate and the thermal cloud in kinetic theories \cite{Proukakis_Thesis,Proukakis_NIST,Proukakis_JPhysB,ZNG_PRL_2,Bijlsma_Zaremba_Stoof,JILA_Kinetic_Theory_1,Davis_Finite_T_GPE}.\\

\noindent {\em Introduction of an Effective Interaction:} 
The usual `naive' justification for the Gross-Pitaevskii Equation, is to ignore
{\em all} non-condensate fluctuations, while simultaneously approximating the interatomic potential by a delta function, via
$V({\bf r}-{\bf r'})=g \delta({\bf r}-{\bf r'})$, where $g$ a suitable effective interaction strength.
While instructive, this is misleading, because
$V({\bf r}-{\bf r'})$ corresponds to the {\em actual} interatomic potential, whereas $g$ is the interaction strength for a complete effective binary s-wave scattering contact interaction in vacuum \cite{Fetter_Walecka}, which generally proceeds via {\em virtual intermediate} excited states, which may be occupied. This `renormalization' of the coupling constant was first discussed for the homogeneous Bose gas by Beliaev \cite{Beliaev} and generalized to finite temperatures by Popov \cite{Popov}; this problem was recently discussed by various approaches by Stoof {\it et al.} \cite{Stoof_Variational,Stoof_PRL}, Proukakis {\it et al.} \cite{Proukakis_Burnett_Stoof,Proukakis_Royal_Soc,Proukakis_GHFB}, Shi and Griffin \cite{Shi_Griffin} and Morgan \cite{Morgan_JPhysB}. We follow the notation of Proukakis {\it et al.} \cite{Proukakis_NIST} and investigate the condensate evolution at zero temperature, where there is a negligible non-condensate population, such that  $\langle \hat{c}_{i}^{\dag} \hat{c}_{j} \rangle \approx \langle \hat{c}_{i}^{\dag} \hat{c}_{j} \hat{c}_{k} \rangle \approx 0$.
Apart from pure mean field effects, $z_{i}^{*}z_{j}z_{k}$, the equation for the mean amplitude $z_{n}$ 
still contains a contribution from the anomalous average.
The latter can be separated via $\langle \hat{c}_{j} \hat{c}_{k} \rangle = \langle \hat{c}_{j} \hat{c}_{k} \rangle_{0} + \delta \langle \hat{c}_{j} \hat{c}_{k} \rangle$
 into a slowly-varying contribution over low-lying states,  $\langle \hat{c}_{j} \hat{c}_{k} \rangle_{0}$, and a contribution over high-lying states,  
$ \delta \langle \hat{c}_{j} \hat{c}_{k} \rangle$, which varies {\em within} a collisional timescale.
By using the corresponding equation of motion for $ \delta \langle \hat{c}_{j} \hat{c}_{k} \rangle$ to adiabatically eliminate this quantity from the 
zero-temperature limit of Eq. (\ref{zn}), the latter takes the form \cite{Proukakis_Thesis,Proukakis_Burnett_Stoof}
\begin{equation}
i \hbar \frac{d z_{n}}{dt} = \hbar \omega_{n} z_{n} + \sum_{ijk} \langle ni | \hat{T}^{\rm 2B} | jk \rangle
z_{i}^{*} \left[ z_{j} z_{k} +  \langle \hat{c}_{j} \hat{c}_{k} \rangle_{0} \right],
\end{equation} 
where $\hat{T}^{\rm 2B}$ represents an effective two-body interaction term obeying a Lippmann-Schwinger relation \cite{Proukakis_Thesis}
expressing the complete repeated scattering amplitude for two particles in an empty trap.
This procedure corresponds to a diagrammatic resummation of a selected class of diagrams \cite{Proukakis_Thesis,Stoof_Variational,Shi_Griffin}.
This effective interaction is strictly only valid when used to describe the coupling between low momentum states \cite{Proukakis_Thesis,Davis_Finite_T_GPE,Morgan_JPhysB,Burnett_Lecture_Notes}, and cannot be used as a real potential acting between the atoms. 
Ignoring all energy and momentum dependence,
this `renormalized' interaction can now be approximated by an effective contact (pseudo)potential
$\hat{T}^{\rm 2B}({\bf r}, {\bf r'})=g \delta ({\bf r}-{\bf r'})$, where $g=4 \pi \hbar^{2} a/m$, and $a$ is the s-wave scattering length  \cite{Fetter_Walecka,Huang_Yang}.
The implicit assumption of a constant, momentum-independent, scattering amplitude must be used
cautiously, and is valid only up to some maximum momentum $k_{\rm max} \propto 1/a$ \cite{Popov}.
The anomalous contribution $ \langle \hat{c}_{j} \hat{c}_{k} \rangle_{0}$ accounts for many-body effects which are typically neglected in dilute weakly-interacting gases (see Sec. \ref{Gen_MF}).
Re-expressing the mean-value equation in terms of the condensate mean field $\Psi({\bf r},t)$, we obtain the Gross-Pitaevskii Equation at $T=0$,
\begin{equation}
i \hbar \frac{\partial}{\partial t} \Psi({\bf r},t) = \left[ \hat{H}_{0} 
+ g \left| \Psi({\bf r},t) \right|^{2} \right]  \Psi({\bf r},t) = \hat{H}_{\rm GP} \Psi({\bf r},t) \label{GPE},
\end{equation}


\noindent {\em Number-Conserving Treatment:}
The above treatment assumes the existence of a mean field, and thus relies 
on spontaneous symmetry breaking. While convenient, this notion is somewhat ill-defined for a finite trapped gas with a fixed atom number $N$, since the condensate mean field 
corresponds to the ensemble average over number states $|N\rangle$, with $ \langle N | \hat{\Psi}({\bf r},t) | N \rangle = \Psi({\bf r},t) \langle N | N-1 \rangle = 0$. 
The usual justification 
is that for a system with a 
large number of particles, 
$N \pm 1 \approx N$ with accuracy $(1/N)$, 
essentially assuming that the condensate operators commute \cite{Bogoliubov}.
Such arguments are 
valid in the grand canonical ensemble, where the atom number is not fixed, and one 
minimizes the shifted hamiltonian $( \hat{H} - \mu \hat{N} )$, where $\mu$ the chemical potential and $\hat{N}$ the number operator. 
Also, no difficulties arise if, by analogy to laser theory, we treat the 
condensate  as a coherent state $| \alpha \rangle$, i.e. a superposition of number states for which $\langle \alpha | \alpha \rangle \neq 0$ by construction.

Gardiner argued that the apparent non-conservation of particle numbers arises from an inappropriate definition of the quasi-particle operators. This led to the development of a $U(1)$-symmetry-preserving approach, relying on the separation of the `particle' and `phonon' concepts \cite{Gardiner_Number_Conserving}.
Following Gardiner's notation, 
$\hat{\Psi}({\bf r}) = \hat{A} [ \xi({\bf r}) + (1/\sqrt{N}) \hat{\chi}({\bf r})]$
where $\xi({\bf r})$ is the condensate wavefunction obeying the usual GPE, $\hat{\chi}({\bf r})$ is a phonon field operator, and $\hat{A}$ is a generalized annihilation operator ensuring total particle number conservation. 
This approach
sets the correct basis for further treatment \cite{QK_PRL_1}, with an equivalent approach formulated by Castin and Dum \cite{Castin_Dum} and extended to finite temperatures by Morgan \cite{Morgan_JPhysB}.

The remaining sections generalize the 
above treatment to finite temperatures.
Crudely speaking, one can divide such generalized treatments into those that rely on a suitably truncated hierarchy of coupled equations of motion for generalized mean fields (Secs. \ref{Gen_MF} \& \ref{Gen_MF_t}), and those in which the lowest modes of the system, which include both condensate and quasi-particles, are described in a unified manner by appropriate probability distribution functions (Sec. \ref{Prob_Function}), giving rise to various stochastic approaches (Sec. \ref{Stochastic}). Sec. \ref{1D} briefly discusses the importance of fluctuations in the phase of the condensate, relevant in low dimensional systems.

\section{Generalized Mean Fields: Static Thermal Cloud \label{Gen_MF}}

From Eq. (3), it is clear that the condensate mean field $\Psi({\bf r},t)$ is coupled to `higher order' mean fields (normal and anomalous averages).
Their corresponding  equations of motion are, in turn,  coupled to yet higher order correlations of the non-condensate operators, and so forth.
A closed set of equations requires a consistent set of decoupling approximations, thereby uniquely defining a set of generalized mean fields which should accurately determine the static and dynamic properties of the system
\cite{Kirkpatrick_Dorfman_1}.
Equations of this form are well-known in the literature and can be derived by variational  \cite{Stoof_Variational,OURREVIEW}, or diagrammatic \cite{Shi_Griffin} methods. A description in terms of the three lowest order averages $\langle \hat{\Psi} \rangle$, $\langle \hat{\psi}'^{\dag} \hat{\psi}' \rangle$ and  $\langle \hat{\psi}' \hat{\psi}' \rangle$ is known as the Hartree-Fock-Bogoliubov (HFB) theory
 (see below) \cite{HFB_Early}.\\ 

\noindent {\em Mean Field Approximation:}
To first approximation,
one can ignore  {\em particle exchange} between condensate and thermal cloud (i.e. the triplets),
and allow only {\em mean-field coupling} between them. 
Such a theory arises upon reducing the hamiltonian of Eq. (\ref{H}) to quadratic form via the `mean field' approximations \cite{Proukakis_Thesis,Griffin_HFB,Morgan_JPhysB}
$\hat{\psi}'^{\dag} \hat{\psi}' \hat{\psi}' \approx 2 \langle \hat{\psi}'^{\dag} \hat{\psi}' \rangle \hat{\psi}' 
+ \hat{\psi}'^{\dag} \langle \hat{\psi}' \hat{\psi}' \rangle$
and
$\hat{\psi}'^{\dag} \hat{\psi}'^{\dag} \hat{\psi}' \hat{\psi}' \approx 
4 \langle \hat{\psi}'^{\dag} \hat{\psi}' \rangle \hat{\psi}'^{\dag} \hat{\psi}' 
+ \langle \hat{\psi}'^{\dag} \hat{\psi}'^{\dag} \rangle \hat{\psi}' \hat{\psi}'
+ \langle \hat{\psi}' \hat{\psi}' \rangle \hat{\psi}'^{\dag} \hat{\psi}'^{\dag}
- 2 \langle \hat{\psi}'^{\dag} \hat{\psi}' \rangle \langle \hat{\psi}'^{\dag} \hat{\psi}' \rangle
- \langle \hat{\psi}' \hat{\psi}' \rangle \langle \hat{\psi}'^{\dag} \hat{\psi}'^{\dag} \rangle$
the latter justified from Wick's theorem \cite{Fetter_Walecka}.
This leads to an effective quadratic hamiltonian in which single-particle energies are shifted by the added mean field potentials. Such a hamiltonian
can be diagonalized by the Bogoliubov transformation
$\psi'({\bf r}) = \sum_{i} \phi_{i}({\bf r}) \hat{c}_{i} = \sum_{j} [ u_{j}({\bf r}) \hat{\beta}_{j} - v_{j}^{*}({\bf r}) \hat{\beta}_{j}^{\dag} ]$,
mapping single-particle non-condensate operators $\hat{c}_{i}$ into mixtures of `quasi-particle' operators $\hat{\beta}_{j}$ and $\hat{\beta}_{j}^{\dag}$, 
%
This leads to a generalized GPE 
\begin{equation}
i \hbar \frac{\partial}{\partial t} \Psi({\bf r},t) = \left[ \hat{H}_{0}
+ g  \left[ |\Psi({\bf r},t)|^{2} + 2 n'({\bf r},t) \right] \right] \Psi({\bf r},t)
+ \tilde{m}_{0}({\bf r},t) \Psi^{*}({\bf r},t),
\end{equation}
which, compared to the GPE of Eq. (\ref{GPE})
contains additional contributions from the {\em mean field coupling} of the
condensate $\Psi$ to the non-condensate $n'$ (with the factor of two arising due to direct and exchange contributions \cite{Fetter_Walecka}) 
and the anomalous average $\tilde{m}_{0}$.
The above equation is
coupled to finite temperature `Bogoliubov-de Gennes' equations for the excitation
amplitudes $u_{j}({\bf r})$, $v_{j}({\bf r})$. 
Such a closed set of equations (see Eqs. (\ref{BdG1})-(\ref{BdG2}))
is known as the HFB theory \cite{Griffin_HFB,Fetter_Walecka,OURREVIEW,HFB_Early}.\\
\noindent {\em Many-Body Effects:}
We argued that the two-body effective interaction, $\hat{T}^{\rm 2B}$, arises via adiabatic elimination of rapidly-varying anomalous averages.
Such an effective interaction describes the collisional amplitude in vacuum, whereas in trapped gases collisions take place within a medium of other condensed and excited particles. This
leads to 
important modifications in the description of atom-atom interactions \cite{Stoof_Variational,Proukakis_Burnett_Stoof,Proukakis_Royal_Soc,Shi_Griffin}
which can again be incorporated into a more general effective interaction, the many-body T-matrix, $\hat{T}^{\rm MB}$.
This 
can be approximated (in the zero-energy, zero-momentum limit) by a contact potential with a
position-dependent amplitude 
$g_{\rm eff}({\bf r}) = g [ 1 + \tilde{m}_{0}({\bf r})/\psi^{2}({\bf r}) ]$
\cite{Proukakis_GHFB,Burnett_Lecture_Notes,Hutchinson_GHFB};
here $\tilde{m}_{0}({\bf r})$ denotes the `regularized' position representation of the anomalous average over low-lying states
(see below).
Since $\tilde{m}_{0}({\bf r})$ 
is 
of the same order of magnitude as the non-condensate density $n'({\bf r})$, its effect should, in general, not be ignored \cite{Yukalov_Kleinert}.\\
%
%
%

%
%
\noindent {\em HFB and Related Approaches:}
To first approximation, the non-condensate component is treated as static, 
yielding the static HFB equations, which, written in a slightly more general notation to facilitate subsequent discussion, take the form
\begin{equation}
\left[ \hat{H}_{0} - \mu
+ g_{\rm con}({\bf r}) |\psi({\bf r})|^{2} + 2 g_{\rm exc}({\bf r}) n'({\bf r}) \right] \psi({\bf r}) = 0 \label{BdG1}
\end{equation}
\begin{eqnarray}
\left( \begin{array}{c} \hat{L} \\ -\hat{M}^{*} \end{array} \begin{array}{c} \hat{M} \\ -\hat{L}^{*} \end{array} \right) \left( \begin{array}{c} u_{j} \\ v_{j} \end{array} \right) = \hbar \omega_{j} \left( \begin{array}{c} u_{j} \\  v_{j}  \end{array} \right) \label{BdG2}
\end{eqnarray}
where
$\hat{L}({\bf r}) = \hat{H}_{0} - \mu
+ 2g_{\rm con}({\bf r}) |\psi({\bf r})|^{2} + 2 g_{\rm exc}({\bf r}) n'({\bf r})$
and
$\hat{M} = g_{\rm con}({\bf r}) \psi^{2}({\bf r})$.

This general form distinguishes between the effective interaction for a collision between two condensate atoms, $g_{\rm con}({\bf r})$, from the corresponding one for collisions involving a condensate and a non-condensate atom,  $g_{\rm exc}({\bf r})$. Solving Eqs. (\ref{BdG1})-(\ref{BdG2}) yields the `Bogoliubov functions' $u_{j}({\bf r})$ and $v_{j}({\bf r})$, in terms of which we can directly obtain the time-independent normal 
$n'({\bf r}) = \sum_{j} ( |u_{j}({\bf r})|^{2} + |v_{j}({\bf r})|^{2} ) N_{\rm BE}(\varepsilon_{j})
+ |v_{j}({\bf r})|^{2} $
and anomalous
$\tilde{m}_{0}({\bf r}) = \sum_{j} u_{j}({\bf r}) v_{j}^{*}({\bf r}) [ 2 N_{\rm BE}(\varepsilon_{j}) + 1 ] 
- \tilde{m}_{\rm UV}({\bf r})$ 
averages. 
Here $N_{\rm BE}(\varepsilon_{j}) = \left[ {\rm exp} \left( \beta \hbar \omega_{j} \right) - 1 \right]^{-1}$ is the Bose-Einstein distribution with $\beta = 1/ k_{B}T$.
The term $\tilde{m}_{\rm UV}({\bf r})$ is required to ensure the regularization of the anomalous average by  subtracting off the ultraviolet divergences arising in the homogeneous limit
due to the use of a momentum-independent contact interaction.

The standard HFB limit is based on $g_{\rm con}({\bf r}) = g_{\rm exc}({\bf r}) = g$ everywhere
{\em except} in Eq. (\ref{BdG1}) where $g_{\rm exc}({\bf r}) = g$ still, but $g_{\rm con}({\bf r}) = g_{\rm eff}({\bf r})$.
This inconsistent treatment of effective interactions leads to the appearance of
a gap in the homogeneous single-particle excitation spectrum at zero momentum, in direct contrast to the Goldstone theorem 
which guarantees a gapless spectrum in the presence of $U(1)$ symmetry-breaking \cite{Goldstone}, and the Hugenholtz-Pines theorem \cite{Hugenholtz_Pines}. As a result, the full HFB theory should not be used to calculate excitation frequencies.
One common way to overcome this problem, is to completely ignore the 
{\em additional} influence of the medium on collisions of trapped atoms,
i.e. to set $\tilde{m}_{0}({\bf r}) = 0$ \cite{Griffin_HFB,Shi_Griffin}. This led to the first quantitative predictions of excitation frequencies of trapped Bose gases at low temperatures \cite{Hutchinson_Excitations}, in agreement with experiments. 
The reduced gapless theory with $\tilde{m}_{0}({\bf r}) = 0$ is often termed `HFB-Popov' approximation
(but see \cite{Yukalov_Kleinert}.)

In order to 
obtain a consistent theory for excitation frequencies, 
but additionally include the modification imposed by the medium on the collisional properties (i.e. $\tilde{m}_{0}({\bf r} \neq 0$)), one may use the above generalized form of the theory, with the effective interaction $g_{\rm con}({\bf r})=g_{\rm eff}({\bf r})$ for the scattering of two atoms in the condensate, and the mean-field coupling between the condensate and non-condensate, $g_{\rm exc}({\bf r})$, approximated  either by $g$, or by $g_{\rm eff}({\bf r})$ consistently throughout Eqs. (\ref{BdG1})-(\ref{BdG2}). These 
correspond to different approximate treatments of the relative momenta of colliding atoms, and give rise to two different generalized `gapless' theories, GHFB1 and GHFB2 \cite {Proukakis_GHFB,Burnett_Lecture_Notes}. 
(see also \cite{Olshanii_Pricoupenko}). 
These theories have been applied to excitation frequencies, 
vortices 
and the study of coherence of 2D condensates \cite{Hutchinson_GHFB,GHFB_Vortex}.
Alternative approaches to 
the HFB gap problem appear in \cite{Yukalov_Kleinert,HFB_Tommasini}.
%


\section{Generalized Mean Fields: Dynamic Thermal Cloud \label{Gen_MF_t}}


So far, the non-condensate has been treated as static. Such approaches cannot account for common dissipative process in the condensate, e.g. {\em Landau} damping (coalescence of two excitations into a single one of higher energy) and {\em Beliaev} damping (breaking up of an excitation into two excitations of lower energies), which require a coupled dynamical description of 
both condensate and non-condensate \cite{Pethick_Book}. A complete theoretical picture must also describe collisions between condensate and non-condensate atoms, as well as between two non-condensate atoms. 



\subsection{Time-Dependent Hartree-Fock-Bogoliubov}

This section outlines the work of the author \cite{Proukakis_Thesis} as finalized in \cite{Proukakis_JPhysB}.
The exact equation of motion for the non-condensate $\langle \hat{c}_{i}^{\dag} \hat{c}_{j} \rangle$ depends on averages of up to four fluctuation operators $\hat{c}_{i} = \hat{a}_{i} - z_{i}$.
Using second order perturbation theory, and considering for simplicity a single condensate mean field amplitude, $z_{0}$, coupled to numerous excited levels of populations $n_{i}$, one obtains 
%
%
coupled equations governing their evolution.
The equation for $dn_{i}/dt$ includes, among other terms, the following contribution accounting for collisions between two non-condensate atoms
\begin{eqnarray}
\left( \frac{4 \pi}{\hbar^{2}} \right) \sum_{rms} 
\left| T_{rsmi}^{\rm 2B} \right|^{2}
\left[ (n_{i}+1) (n_{m}+1) n_{r} n_{s} - n_{i}n_{m} (n_{r}+1) (n_{s}+1) \right]
\delta ( \omega_{rsmi} )
\label{one}
\end{eqnarray}
where $\omega_{rsmi}=\left( \omega_{r}+\omega_{s} - \omega_{m} - \omega_{i} \right)$. 
We also obtain the following {\em additional} contribution to the thermal cloud dynamics $dn_{i}/dt$ due to the presence of a condensate
\begin{eqnarray}
\left( \frac{4 \pi}{\hbar^{2}} \right) |z_{0}|^{2} \sum_{rs} 
\left| T_{rs0i}^{\rm 2B} \right|^{2} 
\left[ (n_{i}+1) n_{r} n_{s} - n_{i} (n_{r}+1) (n_{s}+1) \right] 
\delta ( \omega_{rs0i} )
\label{two}.
\end{eqnarray}
This term is proportional to $|z_{0}|^{2}$ and describes the change in the non-condensate population of level $i$ induced by the transfer of a single atom to, or from the condensate 
(level $0$).
One must also consider the corresponding change in the condensate population associated with such processes.
Adiabatic elimination of the 
triplet correlation $\langle \hat{c}_{i}^{\dag} \hat{c}_{j} \hat{c}_{k} \rangle$ 
in Eq. (\ref{zn})
leads
to \cite{Proukakis_Thesis} 
\begin{eqnarray}
i \hbar \frac{d z_{0}}{dt} & = & \hbar \omega_{0} z_{0} 
+ \left( T_{0000}^{\rm 2B} |z_{0}|^{2} + 2 \sum_{i} T_{0ii0}^{\rm 2B} n_{i} \right) z_{0} \label{three}
\\
& + & i \left( \frac{2 \pi}{\hbar} \right) z_{0} \sum_{irs} \left|T_{rs0i}^{\rm 2B} \right|^{2} 
\left[ (n_{i}+1)n_{r}n_{s} - n_{i} (n_{r}+1) (n_{s}+1) \right] 
\delta \left( \omega_{rs0i} \right) \nonumber
\end{eqnarray}
which clearly contains both energy shifts due to mean fields and damping.
The equations given above contain the essential {\em key elements} 
for a consistent kinetic theory for a partially-Bose-condensed gas.
These reveal that while the scattering of a particle into an excited state $i$ is `bosonically enhanced' by the factor $(n_{i}+1)$, the corresponding scattering into the condensate 
does not feature spontaneous growth 
(i.e. one simply obtains $|z_{0}|^{2}$  and $z_{0}$ in Eqs. (\ref{two}) and (\ref{three}) respectively).

The discussion thus far relied on a perturbative expansion beyond
a single-particle basis of {\em bare} (undressed) particles, and has led to dressing of the energies from the mean fields of the condensate, $|z_{0}|^{2}$, and the non-condensate, $n_{i}$. Such basis choice is, however, not very suitable for typical systems containing a condensate spanning numerous trap eigenstates. 
While one could transform the above equations to a quasi-particle basis, this leads to rather involved expressions \cite{Wachter_MPhil,Imamovic_Griffin}. An alternative, but {\em equivalent}, approach  is to {\em first} incorporate all relevant HFB mean field effects into the basis, i.e. shift from bare single-particle energies $\hbar \omega_{n}=\langle n | \hat{H}_{0} | n \rangle$ to dressed quasi-particle energies $\hbar \omega_{n}^{\rm HFB}=\langle n | \hat{H}_{\rm HFB} | n \rangle$ where $\hat{H}_{\rm HFB}$ is the reduced quadratic hamiltonian with mean field effects included \cite{Fetter_Walecka}, and {\em subsequently} perform second order perturbation theory beyond this dressed basis. In this case, the perturbative terms arise from the difference between the {\em exact} expressions of products of fluctuation operators, and their corresponding HFB mean field decompositions. For example, one part of the perturbative hamiltonian arises from the `beyond-mean-field' quantity
$\langle \hat{\psi}'^{\dag} \hat{\psi}'^{\dag} \hat{\psi}' \hat{\psi}' \rangle - \left(
2 \langle \hat{\psi}'^{\dag} \hat{\psi}' \rangle^{2} + 
\langle \hat{\psi}'^{\dag} \hat{\psi}'^{\dag} \rangle \langle \hat{\psi}' \hat{\psi}' \rangle\right) \label{Perturbative}$
as suggested  by Morgan \cite{Morgan_JPhysB} and Zaremba, Nikuni and Griffin \cite{ZNG}.
Explicitly performing this (see \cite{Proukakis_JPhysB} for details) yields a set of coupled time-dependent non-markovian equations for the mean field amplitudes $z$, $\langle \hat{c}^{\dag} \hat{c} \rangle$ and
$\langle \hat{c} \hat{c} \rangle$, corresponding to $\Psi$, $n'$ and $\tilde{m}_{0}$.
These equations describe all atomic collisions, including not only collisions within the condensate, but also particle exchange between condensate and non-condensate and collisions of two thermal atoms.
This theory was first derived by Walser et al. \cite{JILA_Kinetic_Theory_1,JILA_Kinetic_Theory_2}, using slightly different, but entirely equivalent, arguments (based on 
separation of timescales for the evolution of the relevant physical parameters), and
subsequently obtained by the author \cite{Proukakis_JPhysB}.
The reader is referred to 
\cite{Proukakis_JPhysB,JILA_Kinetic_Theory_1,JILA_Kinetic_Theory_2} for details, 
implementation, and relation to other treatments.

The perturbation theory presented here can also be derived, perhaps somewhat more systematically, from the method of non-commutative cumulants, 
in which the cumulants capture the essential correlations in the system by subtracting factorizable contributions from the correlation functions,
as mentioned above, with such an approach already successfully implemeted in diverse contexts 
\cite{Thorsten_1}.\\

\noindent {\em Gapless Second-Order Perturbation Theory:}
To monitor situations close to equilibrium, 
Giorgini \cite{Giorgini_Linear_Response_1}, and Rusch and Burnett \cite{Rusch_Burnett} performed a linear response treatment around static values of the HFB parameters $z$, $\langle \hat{c}^{\dag} \hat{c} \rangle$, and $\langle \hat{c} \hat{c} \rangle$.
This was discussed in detail in a series of papers by Morgan 
\cite{Morgan_JPhysB},
who additionally reconciled the rigorous introduction of an effective interaction with the number-conserving formalism.
Morgan's approach, based on a perturbative treatment beyond the HFB basis,
accounts for the indirect excitation of the condensate via excitation of the thermal cloud, as needed to interpret the excitation frequencies and damping rates at JILA \cite{Excitation_Frequencies_JILA} 
(see also \cite{Stoof_Excitations,Jackson_Zaremba_3}).
The resulting form of the Bogoliubov-de Gennes equations for the excitations are similar to those of Eqs. (\ref{BdG1})-(\ref{BdG2}), but with additional, or modified terms to ensure orthogonality between condensate and non-condensate and number conservation. Such a theory consistently includes 
damping effects, population changes due to quasi-particle collisions and finite size effects.


\subsection{Theory of Zaremba-Nikuni-Griffin}

Parallel to the above developments, Zaremba, Nikuni and Griffin (ZNG) presented the first computationally implementable generalized mean field theory describing the coupled evolution of condensate and thermal cloud \cite{ZNG}.
The starting point is the same as the time-dependent HFB approach.
While the subsequent treatment ignores the anomalous average
(i.e. many-body effects included in \cite{Proukakis_JPhysB,JILA_Kinetic_Theory_1}), this theory made a significant contribution by implementing local energy and momentum conservation \cite{Kirkpatrick_Dorfman_1}. The resulting kinetic equations are valid in both mean-field and hydrodynamic regimes.
%
In this theory, 
the condensate evolves according to
\begin{equation}
i \hbar \frac{\partial}{\partial t} \Psi({\bf r},t) = \left[ \hat{H}_{0} + g \left[ |\Psi({\bf r},t)|^{2} + 2 n'({\bf r},t) \right]
- i R({\bf r},t) \right] \Psi({\bf r},t) \label{FT_GPE}
\end{equation}
with the coupled non-condensate density $n'({\bf r},t)$ described in terms of a Wigner phase-space representation $f({\bf r}, {\bf p},t)$ via
$n'({\bf r},t) = \int d {\bf p}/(2 \pi/\hbar)^{3} f({\bf r}, {\bf p},t)$, 
with $f$ obeying a generalized `Quantum Boltzmann' kinetic equation \cite{Ergodic_1}
\begin{eqnarray}
\frac{ \partial}{\partial t} f({\bf r}, {\bf p}, t) + \frac{ \bf p}{m} \cdot \nabla f({\bf r}, {\bf p},t) - \nabla V_{\rm eff}({\bf r},t) \cdot \nabla_{p} f({\bf r}, {\bf p},t) 
= C_{12}[f] + C_{22}[f] \label{QBE}
\end{eqnarray}
with an effective mean-field potential
$V_{\rm eff}({\bf r},t) = V_{\rm ext}({\bf r},t) + 2g \left[ |\Psi({\bf r},t)|^{2} +  n'({\bf r},t) \right]$.
This equation contains two collisional terms: $C_{22}$ describing collisions between {\em non-condensate} atoms, and $C_{12}$ describing the {\em transfer of an atom} from the non-condensate into the condensate, and its inverse process. These are given by 
\begin{eqnarray}
C_{22}[f] & = &
\frac{4 \pi}{\hbar} g^{2} \int \frac{ d {\bf p}_{2}}{(2 \pi \hbar)^{3}} \int \frac{ d {\bf p}_{3}}{(2 \pi \hbar)^{3}} \int \frac{ d {\bf p}_{4}}{(2 \pi \hbar)^{3}}(2 \pi \hbar)^{3}
\delta \left( {\bf p}+{\bf p}_{2} - {\bf p}_{3} - {\bf p}_{4} \right) \nonumber \\
& & \delta \left( \varepsilon + \varepsilon_{2} - \varepsilon_{3} - \varepsilon_{4} \right)
\left[ (1+f)(1+f_{2})f_{3}f_{4} - f f_{2} (1+f_{3})(1+f_{4}) \right]
\end{eqnarray}
\begin{eqnarray}
C_{12}[f] & = &
\frac{4 \pi}{\hbar} g^{2} |\Psi|^{2} \int \frac{ d {\bf p}_{2}}{(2 \pi \hbar)^{3}} \int \frac{ d {\bf p}_{3}}{(2 \pi \hbar)^{3}} \int \frac{ d {\bf p}_{4}}{(2 \pi \hbar)^{3}}(2 \pi \hbar)^{3}
\delta \left( m{\bf v}_{c}+{\bf p}_{2} - {\bf p}_{3} - {\bf p}_{4} \right) \nonumber \\
& & \delta \left( \varepsilon_{c} + \varepsilon_{2} - \varepsilon_{3} - \varepsilon_{4} 
\right) (2 \pi \hbar)^{3}
\left[ \delta({\bf p}-{\bf p}_{2})-\delta({\bf p}-{\bf p}_{3})-\delta({\bf p}-{\bf p}_{4}) \right] \nonumber \\
& & \times \left[ (1+f_{2})f_{3}f_{4} - f_{2} (1+f_{3})(1+f_{4}) \right],
\end{eqnarray}
where
$f_{i}=f({\bf r}_{i}, {\bf p}_{i},t)$, $\varepsilon_{i} = p_{i}^{2}/2m + V_{\rm eff}({\bf r})$, and $\varepsilon_{c}=\mu_{c}+mv_{c}^{2}/2$ is the local condensate energy.
The condensate growth term, $-iR({\bf r},t)$ in Eq. (\ref{FT_GPE}) arises from `triplet' correlations and is 
defined by
$R({\bf r},t) = \hbar/2|\Psi|^{2} \int d {\bf p}/(2 \pi \hbar)^{3} C_{12}[f]$.

The original numerical implementation of this theory \cite{Bijlsma_Zaremba_Stoof} successfully addressed the issue of condensate formation by relying on
the ergodic approximation \cite{Ergodic_1}, which assumes that equilibration is rapid for atoms with similar energies, thus enabling the phase-space variable $ f_{i}=f({\bf r}_{i}, {\bf p}_{i},t)$  to be expressed only in terms of an energy variable $E({\bf r}_{i}, {\bf p}_{i}, t)$. 
These calculations were subsequently generalized \cite{Jackson_Zaremba_3,Jackson_Zaremba_1} to strongly non-equilibrium regimes by representing the phase-space density by a collection of N discrete `test particles', with collisions between them handled via an appropriate Monte Carlo sampling scheme.
Despite ignoring many-body effects, 
this theory yields remarkable agreement with various experiments, e.g. scissor's mode,
quadrupole excitations 
and transverse breathing modes of elongated condensates \cite{Jackson_Zaremba_1}. Hydrodynamic collective modes 
and vortex nucleation at finite temperatures were also discussed \cite{ZNG_Nikuni_1,ZNG_Vortex}.
The ZNG theory also explained the observed dark soliton decay in the Hannover experiment
\cite{Hannover_Soliton_Exp}, and ascertained the required conditions for the observation of dark soliton oscillations in elongated harmonically-confined BECs \cite{ZNG_Soliton}.
%
%
%
The ZNG theory 
arises as a limiting case of the approaches of Stoof 
and Gardiner, Zoller and co-workers 
discussed next. 




Damping of excitations was also addressed by a different approach by Fedichev, Shlyapnikov and collaborators who studied stability of vortices and dark solitons by considering the scattering of excitations from such macroscopically excited states \cite{Fedichev_Shlyapnikov_1}, with this formalism applied to bright solitons by Sinha {\it et al.} \cite{Brand_Soliton_Decay}.

\section{Kinetic Theories based on Probability Distribution Functions \label{Prob_Function}}

All above theories 
have the common feature of treating the condensate as a separate entity from the thermal cloud, with the former typically identified as the mean value of the Bose field operator $\langle \hat{\Psi}({\bf r},t) \rangle$.
Such perturbative theories 
cannot predict phenomena in the critical regime, e.g. onset of condensation, or the shift in $T_{c}$.
An alternative `class' of approaches was thus developed, 
in which
one treats the low-lying part of the spectrum, including {\em both} the condensate and the low-lying excitations 
as one entity, with the remaining high-lying modes
treated separately, by means of a Quantum Boltzmann Equation.



\subsection{Stoof's Non-Equilibrium Theory}

Stoof developed a unified finite temperature quantum kinetic theory by investigating the evolution of the entire system by means of a Wigner probability distribution $P [ \phi^{*}, \phi; t ]$ \cite{Stoof_PRL,Stoof_JLTP}.
At time $t$, this distribution gives the probability of the system to be in a coherent state $| \phi ({\bf r}) ;t \rangle = {\rm exp} \{ \int d {\bf r} \phi({\bf r}) \hat{\Psi}^{\dag}({\bf r},t) \} |0 \rangle$, where $|0 \rangle$ is the vacuum state.
Using suitably normalized coherent states, $P [ \phi^{*}, \phi; t ] = Tr \left[ \hat{\rho}(t_{0}) |\phi ;t \rangle \langle \phi ; t | \right]$, where $\hat{\rho}(t_{0})$ the initial density matrix of the system prior to the onset of condensation, the determination of the probability function reduces to the calculation of a functional integral 
containing the probabilities $\left| \langle \phi ; t | \phi_{0} ; t_{0} \rangle \right|^{2}$.
This can be evaluated as a `path integral' over all field evolutions from $t$ to $t_{0}$ and back to $t$, leading 
to the Keldysh non-equilibrium formalism. 
Such an approach, enables a systematic consideration of the fluctuations around the mean field, with all correlation functions obtained from moments of the distribution $P [ \phi^{*}, \phi; t ]$.
The 
derivation of this theory is 
given in \cite{Stoof_JLTP}, with a brief 
insightful account presented in \cite{Stoof_Summary_Paper,Stoof_Les_Houches,Stoof_Duine}.
By working with a quantized quadratic effective action, in which interactions are 
`renormalized' to the many-body T-matrix, one obtains a Fokker-Planck equation for the full probability distribution function $P[ \phi^{*}, \phi ; t ]$.  
One then 
`decouples' 
condensate and non-condensate modes, via the substitution
$P \left[ \phi^{*}, \phi; t \right] = P_{0} \left[ \Phi^{*}, \Phi; t \right] P_{1} \left[ \phi'^{*}, \phi'; t \right]$.
This gives rise to a coupled set of equations: firstly, a Fokker-Planck equation 
for the temporal evolution of the probability distribution of the condensate wavefunction in the presence of a thermal cloud \cite{Stoof_JLTP}
\begin{eqnarray}
i \hbar \frac{ \partial}{\partial t} P_{0} \left[ \Phi^{*}, \Phi ; t \right]  = 
& - & \int d{\bf r} \frac{ \delta}{\delta \Phi({\bf r})} \left[ \hat{H}_{0} - \mu(t) - i R({\bf r},t)+g \left| \Phi({\bf r}) \right|^{2} \right] \Phi({\bf r}) P_{0} \left[ \Phi^{*}, \Phi ; t \right] \nonumber \\
& + & \int d{\bf r} \frac{ \delta}{\delta \Phi^{*}({\bf r})} \left[ \hat{H}_{0} - \mu(t) + i R({\bf r},t)+g\left| \Phi({\bf r}) \right|^{2} \right] \Phi^{*}({\bf r}) P_{0} \left[ \Phi^{*}, \Phi ; t \right] \nonumber \\
& -& \frac{1}{2} \int d{\bf r} \frac{ \delta^{2}}{\delta \Phi({\bf r})\delta      \Phi^{*}({\bf r})} \Phi^{*}({\bf r}) \hbar \Sigma^{K}({\bf r},t) P_{0} \left[ \Phi^{*}, \Phi ; t \right] \label{FP}. 
\end{eqnarray}
Here $i R({\bf r},t)$ corresponds to particle exchange between the condensate and non-condensate, while the Keldysh self-energy $\Sigma^{K}({\bf r},t)$ (see Eq. (\ref{Sk})) expresses thermal fluctuations due to incoherent collisions between condensate and non-condensate atoms.
The above 
equation contains both `streaming' and `diffusion' terms; the absence of third-order derivatives 
(present in \cite{Norrie_Ballagh_Gardiner_PRA,TWA_Steel}),
arises from the consistent elimination of selected terms in the effective action, required to avoid double-counting in the many-body approximation.
The non-condensate dynamics is parametrized by introducing a Wigner distribution via 
$N \left( {\bf r}, {\bf p}, t \right) +1/2 = \int d{\bf r'} e^{-i {\bf p} \cdot {\bf r'} } 
\langle \phi' ( {\bf r}+{\bf r'}/2 ) \phi'^{*} ( {\bf r}-{\bf r'}/2 ) \rangle (t)$
with $N \left( {\bf r}, {\bf p}, t \right)$ 
obeying a Quantum Boltzmann Equation.
The formulation of this theory ensures that the fluctuation-dissipation theorem relating $R({\bf r},t)$ and $\Sigma^{K}({\bf r},t)$ is satisfied at equilibrium, thus guaranteeing that the condensate relaxes to the correct equilibrium distribution. 
Although qualitative predictions regarding condensate formation can be made, the solution of this reduced Fokker-Planck equation under general conditions is complicated, because terms appearing within this equation depend implicitly on $\Phi^{(*)}$
via their dependence on energy.
One therefore typically resorts to approximate solutions, by means of the stochastic GPE discussed in Sec. \ref{Stochastic}.
%
%
Related non-equilibrium quantum field theoretic approaches have been recently
discussed in \cite{Ramos_PRL}.

\subsection{The Gardiner-Zoller Quantum Kinetic Master Equation \label{QKME}}

In a series of papers \cite{QK_Theory_1},
Gardiner, Zoller and co-workers used techniques from quantum optics to derive a quantum kinetic master equation for the condensate, coupled to a time-dependent thermal cloud. 
This treatment is based on splitting the physical system into two `energy bands', each of which has its own evolution, but with the two bands additionally 
exchanging both particles and energy: (i) The `Condensate Band', $R_{C}$, is a band of low-lying states which typically include the condensate, and those low-lying non-condensate modes that are heavily affected by the presence of the condensate.
It is formulated within the Bogoliubov number-conserving approximation,
with the basis expansion performed in terms of many-body eigenfunctions. 
This band is fully described by the total number of atoms and the quantum state of the quasi-particles within $R_{C}$. While the atoms are conserved, the quasi-particles are mixtures of phonon states, and their number may change by the absorption, or creation, of a quasi-particle in a single collisional process. (ii) The `Non-Condensate' Band, $R_{NC}$: Atoms with energies higher than some appropriate cut-off energy, $E_{R}$, are described in terms of a wavelet expansion, with particles binned into discrete shells in phase space, on the basis of their position and momentum. 
This is often treated as being in thermal equilibrium, with a local particle phase-space density $f({\bf r}_{i}, {\bf p}_{i}, t)$, temperature $T$, and chemical potential $\mu$.
The master equation for the condensate band, including both condensate and quasi-particles, is thus obtained \cite{QK_Theory_1}, following the standard methodology of quantum optics \cite{QO_Book_1}.
A simplified version 
gives rise
to a rate equation for the mean number, $N_{c}$, of atoms in the condensate \cite{QK_PRL_1}
$dN_{c}/dt = 2 W^{+}(N_{c}) [ ( 1 - {\rm exp}(- \beta \Delta \mu ) N_{c}+ 1 ]$,
where $W^{+}(N_{c}) \approx \left[ 4m (a k T)^{2} / \pi \hbar^{3} \right] e^{2 \beta \mu}$ denotes a growth rate 
depending on the system scattering length $a$, temperature $T$, and chemical potentials of the condensate and non-condensate, 
with $\Delta \mu$ the difference in chemical potential between 
the two subsystems.
This approach led to the first quantitative predictions of condensate growth \cite{QK_PRL_1}, and to good agreement with experiments \cite{QK_PRL_2,MIT_Growth_Experiment}.
Subsequent improved work also took into account the evolution of occupations of lower trap levels via the ergodic quantum Boltzmann equation. 

\section{Stochastic Approaches to Condensate Dynamics \label{Stochastic}}

\subsection{Classical Field Methods}


Although the preceeding discussion indicates severe limitations to the validity of the GPE at finite temperatures, remarkably the GPE 
{\em can} actually be used to model the evolution of {\em any} system described by the Hamiltonian of Eq. (1) {\em provided that} such a system primarily behaves in a classical manner. 
Such classical evolution was first studied in the context of condensation
by Kagan, Svistunov and Shlyapnikov 
\cite{Svistunov_1}. This approximation was also used to study an ideal superfluid approaching equilibrium in \cite{GPE_Superfluid_Damle}, with a qualitative 2D study of evaporative cooling performed in \cite{Marshall_Choi}, and large scale homogeneous simulations 
performed in \cite{Berloff_Svistunov}.
The use of the GPE to study a weakly-interacting trapped gas
stems from the realization that when all relevant modes of the system are highly occupied ($n_{k} \gg 1$), the classical fluctuations of this field are much larger than quantum fluctuations. Such modes can then be fairly accurately represented by a coherent wavefunction, in analogy to the classical description of highly-occupied laser modes. \\


\noindent {\em The Projected Gross-Pitaevskii Equation:}
A classical description can only work for modes up to a certain energy cut-off, since the condition of high occupation will break down for high-lying modes. 
Such a cut-off can be implemented through the use of a projector, $\hat{P}$, onto highly-populated modes, giving rise to 
\begin{equation}
i \hbar \frac{ \partial}{\partial t} \Psi ({\bf r},t) = \hat{H}_{0} \Psi({\bf r},t)+g \hat{P} \left\{ \left| \Psi({\bf r},t) \right|^{2} \Psi({\bf r},t) \right\} + f(\hat{\eta}). \label{PGPE}
\end{equation}
The term $ f( \hat{\eta})$ appearing above is a rather complicated function of the `fluctuation operator' $  \hat{\eta} = ( \hat{1}- \hat{P} ) \hat{\Psi}({\bf r})$ \cite{Davis_Finite_T_GPE,PGPE_PRL}, describing the coupling of the classical region to a heat bath, whose modes are not highly populated and are orthogonal to the condensate.
Ignoring $ f( \hat{\eta})$ leads to the
`Projected Gross-Pitaevskii Equation' (PGPE) introduced by Davis, Morgan and Burnett \cite{PGPE_PRL}.
This is similar to the usual GPE, Eq. (\ref{GPE}), apart from the presence of the projection operator, $\hat{P}$, which ensures that all computed quantities remain within the classical region.
The projector used in this theory is diagonal in the single-particle basis of the hamiltonian  and
is defined by
$\hat{P} \{ F({\bf r,t}) \} = \sum_{n \epsilon C} \phi_{n}({\bf r}) \int d^{3}{\bf r} \phi_{n}^{*} ({\bf r'}) F({\bf r' },t)$
where $\phi_{n}({\bf r})$ denotes the $n^{\rm th}$ eigenfunction, 
with $n$ restricted within the coherent (classical) region $C$.
Although the PGPE contains no damping terms, since the quantity $f(\hat{\eta})$ has been dropped, 
the presence of elastic collisions ensures that 
a highly non-equilibrium initial configuration relaxes to an equilibrium distribution of given energy, 
without being particularly sensitive to the initial conditions. Such a distribution can be assigned a temperature, upon noting that the mean occupation will actually be given by the 
classical limit $n(\varepsilon_{i})=\left[ \beta \left( \varepsilon_{i}-\mu \right) \right]^{-1}$ of the Bose-Einstein distribution.
Importantly, the field $\Psi({\bf r},t)$ obtained from the PGPE represents the quantum field of {\em many} low-lying modes, rather than a {\em single} condensate mode.
This theory has been used to investigate the evolution of vorticity in homogeneous systems 
and generalized to trapped systems to discuss the process of evaporative cooling, 
the shift in $T_{c}$ 
and the study of spontaneous vortex-antivortex pair production in quasi-2D gases 
\cite{PGPE_PRA}.
Since the PGPE is non-perturbative,
it can be applied even at the critical region.
%
Related classical field approaches
studying the equilibration of a
suitably randomized initial condition of specified energy
by means of the GPE appear in \cite{Classical_Rzazewski_1}.\\ 

\noindent {\em The Truncated Wigner Approximation:}
The idea of evolving randomized initial conditions via a classical field equation, such as the GPE, can be rigorously justified by consideration of the Wigner quasi-distribution function. 
In general, the equation of motion for the Wigner distribution of a gas of ultracold atoms contains both first and third order derivatives with respect to the atomic field, and the Truncated Wigner Approximation (TWA) amounts to ignoring the latter higher-order derivatives. 
This is justified provided that the number of particles in the system is much larger than the number of accessible modes, such that all modes have an occupation larger than unity.
Thus, the main idea behind this approach is to create 
a set of random classical fields which accurately sample the Wigner distribution function of the density operator of the system, and evolve these via the Gross-Pitaevskii equation. 
%
Despite the deterministic nature  of this equation, effects of quantum noise are actually maintained in an approximate manner in the initial state, in the form of mode amplitude fluctuations \cite{Norrie_Ballagh_Gardiner_PRA}, which should be appropriately sampled. 
%
%
The first numerical implementation of the TWA was performed in \cite{TWA_Steel}, with appropriate sampling techniques 
analysed in 
\cite{TWA_Castin_PRL}. 
The TWA was recently used to study colliding condensates \cite{Norrie_Ballagh_Gardiner_PRA}, condensate reflection from a steep barrier, 
three-body recombination processes, 
and collapsing condensates 
\cite{Scott_Hutchinson_Gardiner}.
Polkovnikov performed a systematic perturbation theory in quantum fluctuations around the classical system evolution, obtaining the GPE to lowest order and the TWA to next order
\cite{Polkovnikov_TWA_1}.


In addition to the (truncated) Wigner representation, a quantum system described by a master, or Fokker-Planck, equation, can also be studied within the context of the positive-P representation 
\cite{QO_Book_1},
which gives exact results provided that the ensemble averages converge, thus restricting both system size and evolution timescales. The evaporative cooling dynamics were studied by Drummond and Corney \cite{Drummond_Corney}, with significant progress made recently \cite{Deuar_Drummond_xxx}.
This discussion
omits various related techniques, e.g. Monte-Carlo methods.

In our treatment so far, the system relaxes to some equilibrium state, due to noise implemented in the initial conditions only. However, the only way to guarantee that the system will relax to the correct equilibrium is to allow the condensate to be coupled to a thermal cloud,
which provides in an intuitive manner the necessary irreversibility.
One possibility to extend beyond the PGPE or the TWA is to treat the high energy modes as a thermal reservoir, as discussed below.

\subsection{The Stochastic Gross-Pitaevskii Equation \label{Stochastic_GPE}}

The Fokker-Planck equation derived by Stoof \cite{Stoof_JLTP}, Eq. (\ref{FP}), can be mapped directly onto a Langevin equation \cite{QO_Book_1} for the order parameter of the system $\Phi({\bf r},t)$, with multiplicative noise \cite{Stoof_Duine}. This problem can be simplified considerably by assuming that the thermal cloud is sufficiently close to equilibrium, that it can be described by  a Bose distribution 
with a chemical potential $\mu$ and temperature $T$.
In this regime, the thermal cloud acts as a heat bath to the condensate, and the Langevin equation for the condensate takes the form \cite{Stoof_Stochastic}
\begin{equation}
i \hbar \frac{\partial}{\partial t} \Phi({\bf r},t) = \left[ \left( \hat{H}_{GP} - \mu \right)- iR({\bf r},t) \right] \Phi({\bf r},t)
+ \eta({\bf r},t) \label{SGPE}
\end{equation}
where $\hat{H}_{GP}=\hat{H}_{0} + g |\Phi({\bf r},t)|^{2}$ is the usual `Gross-Pitaevskii' hamiltonian.
This Stochastic GPE (SGPE) describes both mean field effects of the condensate and low-lying excited states, {\em and} fluctuations about the mean field,
thus providing valuable information on both diagonal and off-diagonal elements of the one-particle density matrix.
In the `classical' approximation, where $N(\varepsilon_{i})=\left[ \beta (\varepsilon_{i}-\mu) \right]^{-1}$, the dissipative term obeys $i R({\bf r},t) = - (\beta/4)\hbar \Sigma^{K}({\bf r},t) \hat{H}_{GP}$. 
The contribution  $\eta({\bf r},t)$ denotes a `noise term', with Gaussian correlations
$\langle \eta^{*}({\bf r},t) \eta({\bf r'},t') \rangle = i (\hbar^{2}/2) \Sigma^{K}({\bf r},t) \delta (t-t') \delta ({\bf r}-{\bf r'})$.
The equivalence between Eq. (\ref{SGPE}) and 
Eq. (\ref{FP}) comes about because averaging the product $\Phi^{*}({\bf r},t) \Phi({\bf r},t)$ over the different realizations of the noise $\eta({\bf r},t)$ is {\em by construction} equivalent to averaging over the Wigner distribution $P[\Phi^{*}, \Phi; t]$.
The Keldysh self-energy $\Sigma^{K}({\bf r},t)$ describing the strength of the thermal fluctuations due to incoherent collisions between condensate and non-condensate atoms is given by
\begin{eqnarray}
\Sigma^{K}({\bf r},t) & = &
- i \left( \frac{4 \pi}{\hbar} \right) g^{2} \int \frac{ d {\bf p}_{2}}{(2 \pi \hbar)^{3}} \int \frac{ d {\bf p}_{3}}{(2 \pi \hbar)^{3}} \int \frac{ d {\bf p}_{4}}{(2 \pi \hbar)^{3}}(2 \pi \hbar)^{3}
\delta \left( {\bf p}_{2} - {\bf p}_{3} - {\bf p}_{4} \right) 
\nonumber \\
& & \delta \left( \varepsilon_{c}+\varepsilon_{2}-\varepsilon_{3}-\varepsilon_{4} \right)
\times \left[ (1+N_{2})N_{3}N_{4} + N_{2} (1+N_{3})(1+N_{4}) \right] \label{Sk}
\end{eqnarray}
where $N_{i} = N(\varepsilon_{i})$ the Bose distribution for the eliminated part of the gas, with energy $\varepsilon_{i} = ( {\bf p}_{i}^{2} + V_{\rm ext}({\bf r}) + 2 g | \Phi |^{2} )$, and $\varepsilon_{c}$ corresponds to the average local chemical potential.
A simplified form of this SGPE, assuming time-independent occupation numbers in the non-condensate, and thus a time-independent self-energy $\Sigma^{K}({\bf r})$ has been used to study \cite{Stoof_Stochastic} the reversible condensate formation observed experimentally when cycling through the phase transition \cite{MIT_Dimple_Growth}, quasi-condensate growth on an atom chip, 
fluctuations of one-dimensional Bose gases,
and the growth of coherence of an atom laser 
\cite{Proukakis_Growth}.
This theory is also amenable to variational calculations, and such a technique has been used to discuss collisional frequencies and damping rates of collective excitations, growth-collapse cycles in attractive condensates \cite{Stoof_Duine}, and finite temperature dynamics of a single vortex \cite{Stoof_Vortex}. 
%
%

Additionally ignoring the `noise term' $\eta$ in Eq. (\ref{SGPE}), in the `classical' approximation discussed here, the condensate evolution can be cast in the form
\begin{equation}
i \hbar \frac{\partial}{\partial t} \Phi({\bf r},t) = ( 1 - i \gamma) \left( \hat{H}_{GP} - \mu \right) \Phi({\bf r},t), \label{Damped_GPE}
\end{equation}
with a temperature- and position-dependent damping rate $\gamma = i (\beta / 4) \hbar \Sigma^{K}({\bf r})$.
Phenomenological damping of this form 
was originally proposed 
by Pitaevskii \cite{Pitaevskii_Phenomenology}, and first implemented to trapped Bose gases with a constant, position-independent rate $\gamma$ by Choi {\em et al.} \cite{Choi_Phenomenology}.
A related phenomenologically-damped  equation with the factor $(1-i \gamma)$ appearing on the {\em left} hand side of the equation has been used in diverse studies, including vortex lattice growth 
and dark soliton decay 
\cite{Vortex_Lattice_Ueda_1}.
%

A similar 
SGPE has been recently derived by Gardiner {\it et al.} 
\cite{Gardiner_Stochastic_GPE_1} in terms of the Wigner function representation. This combines the ideas of the quantum kinetic theory of Gardiner, Zoller and co-workers 
\cite{QK_Theory_1}
with those of the finite temperature GPE 
\cite{Davis_Finite_T_GPE}. The system is formally split into low- and high-lying modes by means of the projectors $\hat{P}$ and $(1-\hat{P})$, with the high-lying modes treated as a heat bath.  This leads to a master equation for the interaction of a condensate with a fixed bath of non-condensed atoms, which can be
mapped onto a Fokker-Planck equation. 
Subtle differences arise between the approaches of Stoof and Gardiner, in 
the effective interactions (many-body vs. two-body), the elimination of high energy modes, and the introduction and type of noise terms.
When projection operators and noise contributions are ignored, this equation leads 
precisely to Eq. (\ref{Damped_GPE}), which has been used to study rotating condensates and vortex lattice formation \cite{Penckwitt}.

\section{The Role of System Dimensionality \label{1D}}

In a homogeneous system, BEC arises only in 3D (or 2D at $T=0$).
Recent experimental progress has enabled the achievement of weakly-interacting
quasi-2D (quasi-1D) geometries, in which motion in the transverse direction(s) is `frozen out' by tight harmonic confinement, whose typical level spacing $\hbar \omega_{\perp}$
exceeds both thermal and interaction energies.
Although condensation is approximately regained in such reduced dimensionality systems at sufficiently low temperatures, a partially-condensed system, in general, suffers from phase fluctuations which limit the coherence to a region smaller than the `condensate' spatial extent, as experimentally observed \cite{Orsay};
the system is then said to contain a `quasi-condensate' \cite{Popov}.
While the stochastic theories presented earlier can be used to describe such systems \cite{Proukakis_Growth},
any attempt to perform a
mean field analysis of this regime requires consideration of the `condensate operator'
$\hat{\zeta}$ of Eq. (\ref{zet}) \cite{Shlyapnikov_QC}. In general, the Bose field operator should be expressed as
$\hat{\Psi}({\bf r})=\sqrt{n_{0}({\bf r})}e^{i \hat{\chi}({\bf r})}+ \hat{\psi}'({\bf r})$ 
\cite{Stoof_1D_PRL},
where, $n_{0}$ is the quasi-condensate density, and $\hat{\chi}({\bf r})$ an operator accounting for phase fluctuations.
Clearly $n_{0}$ must vanish in the thermodynamic
limit in 1D and 2D ($T \neq 0$) and 
yield
the condensate density in 3D, which occurs when 
the condensate density is identified as ${\rm Lim} \left\{ n_{0} {\rm exp} \{- (1/2) [ \hat{\chi}({\bf r})-\hat{\chi}({\bf r'}) ]^{2} \} \right\}$ when $({\bf r}-{\bf r}') \rightarrow \infty$. 
%
%
This approach 
provides {\em ab initio} predictions for the equilibrium
properties of {\em weakly-interacting} trapped gases in {\em all} dimensions, 
in agreement with the stochastic GPE \cite{Proukakis_Growth}.\\

As soon as phase fluctuations come into play, the GPE fails to predict the behaviour of the condensate, or its excitations.
Therefore, all predictions of the 1D or 2D GPE presented in this book are necessarily only valid in appropriately elongated geometries which are, however, {\em sufficiently far} both from the regime of quasi-1D (quasi-2D) behaviour and from $T_{c}$.
Finally, we note that the very rich 1D and 2D strongly-interacting regimes can also not be described by the GPE.
%
As a result, one should generally use the Gross-Pitaevskii Equation with caution when describing the dynamics of macroscopic excitations (solitons, vortices), for
systems of low dimensionality at nonzero temperatures. \\
\noindent {\em Acknowledgments:}
Over the past few years I have had the pleasure of numerous extended discussions and collaborations with Keith Burnett and Henk Stoof. I would also like to thank David Hutchinson and Brian Jackson for detailed comments on this manuscript, and
Usama Al Khawaja, Rob Ballagh, Matt Davis, Reinhold Walser,  and Eugene Zaremba for interesting discussions.





\end{document}